\documentclass{article}
\usepackage[preprint]{gpsmdms_2022}
\usepackage{amsmath, amssymb, bm, graphicx}

\DeclareMathOperator*{\argmin}{arg\,min}
\DeclareMathOperator*{\argmax}{arg\,max}
\newcommand{\indep}{\mathop{\perp\!\!\!\!\perp}}

\title{Bayesian Sequential Experimental Design for a Partially Linear Model with a Gaussian Process Prior}
\author{Shunsuke Horii\\
Waseda University\\
\texttt{s.horii@waseda.jp}}
\date{September 2022}

\begin{document}

\maketitle

\begin{abstract}
    We study the problem of sequential experimental design to estimate the parametric component of a partially linear model with a Gaussian process prior.
    We consider an active learning setting where an experimenter adaptively decides which data to collect to achieve their goal efficiently.
    The experimenter's goals may vary, such as reducing the classification error probability or improving the accuracy of estimating the parameters of the data generating process.
    This study aims to improve the accuracy of estimating the parametric component of a partially linear model.
    Under some assumptions, the parametric component of a partially linear model can be regarded as a causal parameter, the average treatment effect (ATE) or the average causal effect (ACE).
    We propose a Bayesian sequential experimental design algorithm for a partially linear model with a Gaussian process prior, which is also considered as a sequential experimental design tailored to the estimation of ATE or ACE.
    We show the effectiveness of the proposed method through numerical experiments based on synthetic and semi-synthetic data.
\end{abstract}

\section{Introduction}
Statistical inference and machine learning can be described as the problem of deriving some decisions from data.
For example, the goal of many statistical inference problems is to accurately estimate some feature of the distribution of the data-generating mechanism, while the purpose of prediction problems is to construct a predictor with good prediction performance.
Many statistical inference methods and machine learning algorithms first gather a large amount of data randomly sampled from the underlying data-generating distribution. 
Then they output an estimation of an estimand or a predictor.
However, in some cases, we have limited resources for collecting such data.
In such cases, it is valuable to determine how we can use these resources as much as possible.
In statistics, sequential experimental design studies ways to sample adaptively to improve estimation efficiency \cite{seeger2007bayesian, ryan2016review, chaloner1995bayesian}. 
In machine learning, the study of sequentially obtaining labeled data from unlabeled data is called active learning \cite{settles2009active, ren2021survey, mackay1992information}.
The study on a sequential design strategy for global optimization of black-box functions is also called Bayesian optimization \cite{NIPS2014_069d3bb0}.
In a sequential experimental design, the sample point to be obtained is determined adaptively so that the estimation accuracy of the estimand is as good as possible.
A typical estimand is the regression coefficients of a linear regression model \cite{seeger2007bayesian, chaloner1995bayesian}.
In this study, we consider a sequential experimental design in which estimand is a causal parameter of the data-generating distribution.
The problem of causal inference, which is the estimation of the effect of changing the value of a treatment variable on the outcome variable, is one of the most important problems in data science.
Although there are other studies of sequential experimental design for causal inference, these studies aim at estimating the structure of the causal graph \cite{he2008active, HAUSER2014926}.
On the other hand, our study aims at estimating a causal parameter known as the average treatment effect (ATE) or the average causal effect (ACE).

In this study, the relationship between variables is modeled by a partially linear model.
The study of a partially linear model dates back to Engle et al \cite{engle1986semiparametric}. 
The partially linear model combines the linear and nonparametric regression models.
Under some assumptions, the linear component of the model can be regarded as a causal parameter, and Double/Debiased machine learning (DML) \cite{10.1111/ectj.12097}, which has been actively studied in recent years, can be seen as one of the estimation algorithms of it.
In this study, we consider the partially linear model proposed in \cite{choi2015partially}, which assumes a Gaussian process prior distribution for the nonlinear component.
We define an information gain based on the posterior distribution of the causal parameter.
We can efficiently estimate the causal parameter by adaptively obtaining data with the largest information gain.

\section{Partially Linear Model with Gaussian Process Prior}
This study investigates the causal effect of the binary treatment variable $T\in\left\{0, 1\right\}$ on the outcome variable $Y\in\mathbb{R}$.
In addition to $T, Y$, we assume that there are some other covariates $\bm{X}\in\mathbb{R}^{d}$.
We model the relation between $T, Y, \bm{X}$ by the following model.
\begin{align}
    Y_{i}=\theta T_{i}+f(\bm{X}_{i})+\varepsilon_{i},\qquad \mathrm{E}\left[\varepsilon_{i}|T_{i},\bm{X}_{i}\right]=0, \label{eq:model}
\end{align}
where $f(\cdot)$ is an unknown nonlinear function from $\mathbb{R}^{d}$ to $\mathbb{R}$.
The model (\ref{eq:model}) is an example of partially linear models since the right-hand side is linear in $T_{i}$ and nonlinear in $\bm{X}_{i}$.
Furthermore, we assume that the nonlinear function $f$ follows a Gaussian process prior $\mathcal{GP}(\bm{0}, C(\cdot, \cdot; \bm{\omega}))$ defined by the covariance function $C(\cdot, \cdot, \bm{\omega})$ with parameter $\bm{\omega}$, and the error term $\varepsilon_{i}$ is i.i.d. with a Gaussian distribution $\mathcal{N}(0, s_{\varepsilon}^{-1})$.
This is an instance of the model proposed in \cite{choi2015partially}.

We are interested in the estimation of $\theta$ since it can be considered as a causal parameter $\mathrm{E}[Y^{(1)}-Y^{(0)}]$ or $\mathrm{E}_{\mathrm{do}(T=1)}[Y]-\mathrm{E}_{\mathrm{do}(T=0)}[Y]$ under some assumptions.
See the appendix \ref{sec:appendix_a} for the relation between $\theta$ and causal parameters.
Among various possible methods for estimating $\theta$, we consider an estimation method based on the posterior distribution of $\theta$.
For this purpose, we assume that $\theta$ is a random variable with a Gaussian prior $\mathcal{N}(\mu_{\theta,0}, s_{\theta,0}^{-1})$.
Given the observations $\bm{t}=[t_{1},\ldots,t_{n}]^{\top}$, $\bm{y}=[y_{1},\ldots,y_{n}]^{\top}$, and $\bm{X}=[\bm{x}_{1}^{\top},\ldots,\bm{x}_{n}^{\top}]^{\top}$, the posterior distribution of $\theta$ is given by
\begin{align}
    p(\theta|\bm{t},\bm{y},\bm{X})&=\mathcal{N}(\mu_{\theta,n}, s_{\theta,n}^{-1}),\\
    \mu_{\theta,n}&=s_{\theta,n}(s_{\theta,0}\mu_{\theta,0}+\bm{t}^{\top}(s_{\varepsilon}^{-1}\bm{I}_{n}+\bm{\Phi})^{-1}\bm{y}),\label{posterior_mean}\\
    s_{\theta,n}&=s_{\theta,0}+\bm{t}^{\top}(s_{\varepsilon}^{-1}\bm{I}_{n}+\bm{\Phi})^{-1}\bm{t},\label{posterior_precision}
\end{align}
where $\bm{I}_{n}$ is the $n$-dimensional identity matrix and $\bm{\Phi}$ is the covariance matrix of $[f(\bm{x}_{1}),\ldots,f(\bm{x}_{n})]^{\top}$ given by
\begin{align}
    \bm{\Phi}=\left[
    \begin{array}{cccc}
         C(\bm{x}_{1},\bm{x}_{1};\bm{\omega})& C(\bm{x}_{1},\bm{x}_{2};\bm{\omega}) & \cdots & C(\bm{x}_{1},\bm{x}_{n};\bm{\omega})\\
         C(\bm{x}_{2},\bm{x}_{1};\bm{\omega})& C(\bm{x}_{2},\bm{x}_{2};\bm{\omega}) & \cdots & C(\bm{x}_{2},\bm{x}_{n};\bm{\omega})\\
         \vdots & \vdots & \ddots & \vdots\\
         C(\bm{x}_{n},\bm{x}_{1};\bm{\omega})& C(\bm{x}_{n},\bm{x}_{2};\bm{\omega}) & \cdots & C(\bm{x}_{n},\bm{x}_{n};\bm{\omega})
    \end{array}
    \right].
\end{align}
The posterior mean $\mu_{\theta,n}$ is the Bayes optimal estimator in the sense that it minimizes the Bayes risk function for the squared error loss.

The model (\ref{eq:model}) has the hyperparameters $\mu_{\theta}, s_{\theta}, \bm{\omega}, s_{\varepsilon}$.
One way to handle them is to assume a prior distribution for them.
In that case, however, the analytic form of the posterior distribution of $\theta$ is no longer available, so we have to rely on numerical computation methods such as Markov Chain Monte Carlo (MCMC) method.
These calculation methods are not the subject of our study, we refer the reader to \cite{choi2015partially} for details.

\section{Sequential Experimental Design based on Information Gain}
In this section, we first describe the problem of the sequential experimental design.
Let $\mathcal{X}_{\mathrm{pool}}$ be the set of $d$-dimensional candidate covariate vectors to sample.
Our problem of sequential experimental design is to determine $\bm{x}_{n+1}\in\mathcal{X}_{\mathrm{pool}}$ and $t_{n+1}\in\left\{0,1\right\}$ given $\bm{t}, \bm{y}$, and $\bm{X}$.
Inspired by the information-theoretic methods for active data collection \cite{seeger2007bayesian, chaloner1995bayesian, mackay1992information, NIPS2014_069d3bb0}, we define the information gain for our problem as follows.
\begin{align}
    g(\bm{x}, t)=\mathrm{H}\left[p(\theta|\bm{t},\bm{y},\bm{X})\right]-\mathrm{E}_{p(y|\bm{t}, \bm{y},\bm{X}, t, \bm{x})}\left[\mathrm{H}\left[p(\theta|\bm{t},\bm{y},\bm{X},t, y, \bm{x})\right]\right],\label{information_gain}
\end{align}
where $\mathrm{H}[p(\theta)]=-\int p(\theta)\ln p(\theta)\mathrm{d}\theta$ represents the differential entropy.
This means that we consider data that reduces the entropy of the posterior distribution of $\theta$ the most informative.
The optimal design is to select $(\bm{x}_{n+1}, t_{n+1})$ that maximizes the information gain, that is, 
\begin{align}
    (\bm{x}_{n+1}, t_{n+1})=\argmax_{\bm{x}\in\mathcal{X}_{\mathrm{pool}},t\in\left\{0, 1\right\}}g(\bm{x}, t).\label{optimal_design_1}
\end{align}
Since the posterior distribution is Gaussian, its entropy is given by
\begin{align}
    \mathrm{H}[p(\theta|\bm{t},\bm{y},\bm{X})]=\frac{1}{2}(1+\ln 2\pi s_{\theta,n}^{-1}).\label{entropy}
\end{align}
Noting that the right-hand side of (\ref{entropy}) does not depend on $\bm{y}$ and the first term in (\ref{information_gain}) does not depend on $(\bm{x},t)$, (\ref{optimal_design_1}) can be written as 
\begin{align}
    (\bm{x}_{n+1}, t_{n+1})=\argmin_{\bm{x}\in\mathcal{X}_{\mathrm{pool}},t\in\left\{0, 1\right\}}\frac{1}{2}\left(1+\ln 2\pi (s_{\theta, n+1}(\bm{x},t))^{-1}\right),
\end{align}
where
\begin{align}
    s_{\theta, n+1}(\bm{x},t)&=s_{\theta,0}+[\bm{t}^{\top}, t](s_{\varepsilon}^{-1}\bm{I}_{n+1}+\bm{\Phi}'(\bm{x}))^{-1}[\bm{t}^{\top}, t]^{\top},\\
    \bm{\Phi}'(\bm{x})&=\left[
    \begin{array}{ccccc}
         C(\bm{x}_{1},\bm{x}_{1};\bm{\omega})& C(\bm{x}_{1},\bm{x}_{2};\bm{\omega}) & \cdots & C(\bm{x}_{1},\bm{x}_{n};\bm{\omega}) & C(\bm{x}_{1},\bm{x};\bm{\omega})\\
         C(\bm{x}_{2},\bm{x}_{1};\bm{\omega})& C(\bm{x}_{2},\bm{x}_{2};\bm{\omega}) & \cdots & C(\bm{x}_{2},\bm{x}_{n};\bm{\omega}) & C(\bm{x}_{2},\bm{x};\bm{\omega})\\
         \vdots & \vdots & \ddots & \vdots & \vdots\\
         C(\bm{x}_{n},\bm{x}_{1};\bm{\omega})& C(\bm{x}_{n},\bm{x}_{2};\bm{\omega}) & \cdots & C(\bm{x}_{n},\bm{x}_{n};\bm{\omega})& C(\bm{x}_{n},\bm{x};\bm{\omega})\\
         C(\bm{x},\bm{x}_{1};\bm{\omega})& C(\bm{x},\bm{x}_{2};\bm{\omega}) & \cdots & C(\bm{x},\bm{x}_{n};\bm{\omega})& C(\bm{x},\bm{x};\bm{\omega})
    \end{array}
    \right].
\end{align}

The proposed optimal design (\ref{optimal_design_1}) is similar to the methods proposed in \cite{seeger2007bayesian, chaloner1995bayesian, mackay1992information,  NIPS2014_069d3bb0}.
However, the objective of those studies is to estimate the function that represents the relationship between inputs and outputs, or the whole data-generating distribution, whereas the objective of our study is to estimate the causal parameter $\theta$.
In our problem, the nonlinear function $f$ is a nuisance parameter, that is, it is noise for estimating $\theta$.
In \cite{van2011targeted}, it is pointed out that plug-in estimation, in which the data-generating distribution is estimated and then the causal parameter is estimated, may not have the desired properties.
They call targeted learning a learning method that is tailored to an estimand such as a causal parameter, so we can call our method a targeted sequential experimental design.

\section{Experiments}
In this section, we illustrate the performance of the optimal design on synthetic data and semi-synthetic data.

\subsection{Experiments on Synthetic Data}
\begin{figure}[t]
  \begin{center}
   \def\@captype{table}
   \begin{minipage}[c]{.48\textwidth}
	\begin{center}
	  \includegraphics[keepaspectratio=true,width=\linewidth]{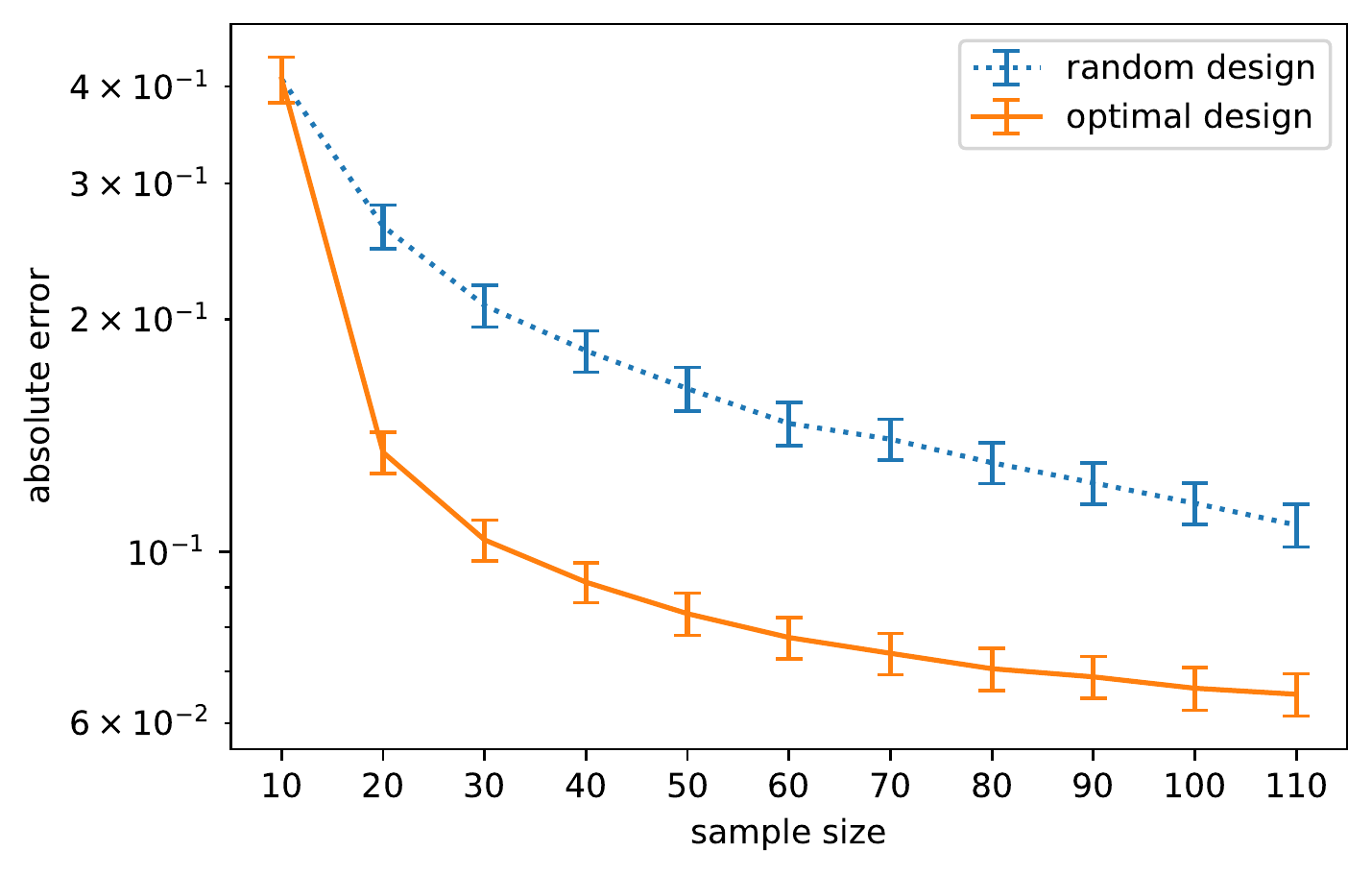}
    \caption{Experimental results on synthetic data. Average$\pm 95\%$ confidence intervals are depicted.}
    \label{fig:synthetic}
	 \end{center}
\end{minipage}  
  \hfill
	\begin{minipage}[c]{.48\textwidth}
	 \begin{center}
	  \includegraphics[keepaspectratio=true,width=\linewidth]{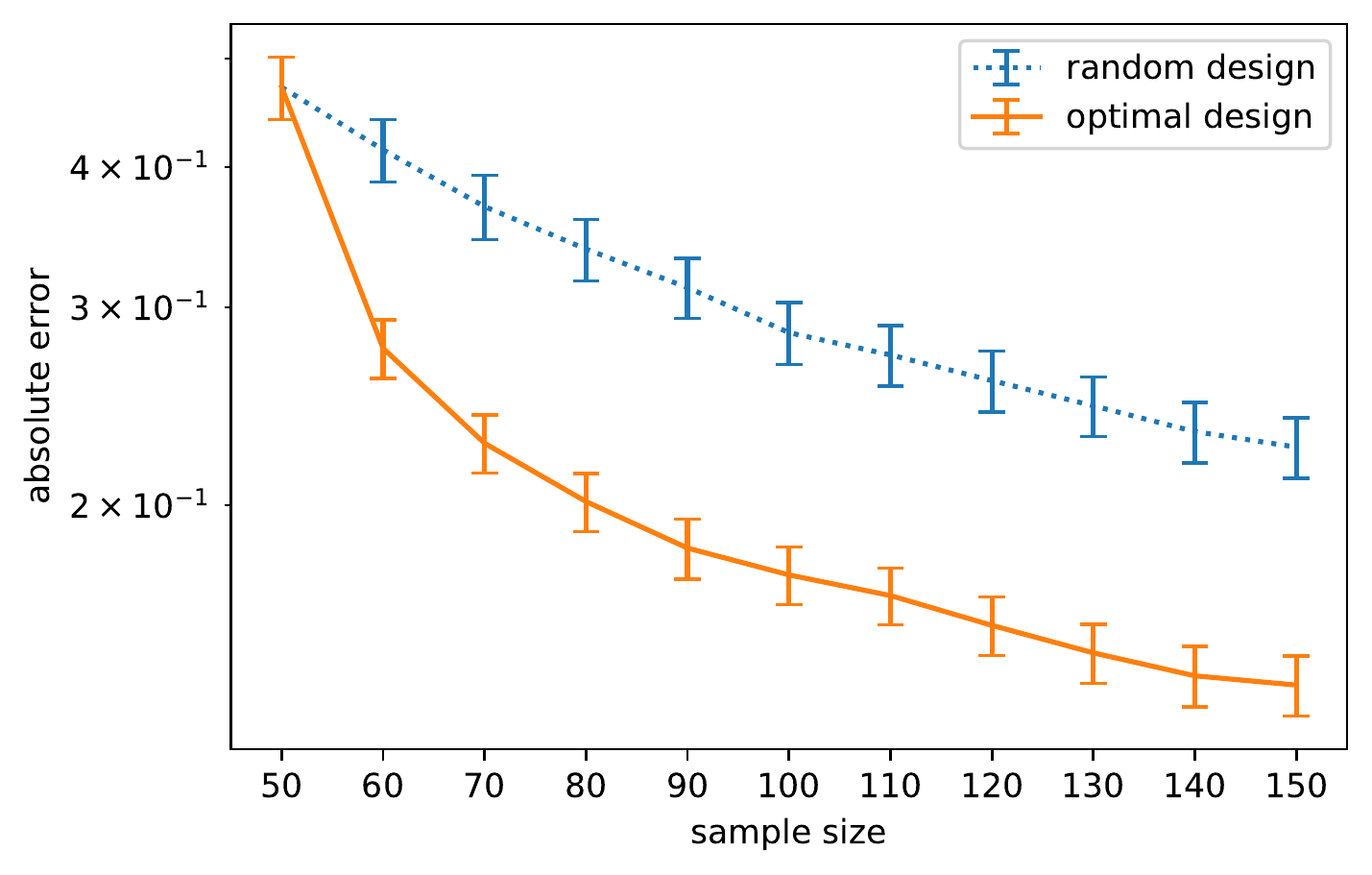}
    \caption{Experimental results on semi-synthetic data. Average$\pm 95\%$ confidence intervals are depicted.}
    \label{fig: semi-synthetic}
	 \end{center}
   \end{minipage}
   \end{center}
 \end{figure}

We illustrate the behavior of the optimal design using synthetic data.
In this experiment, we generate $\theta$ according to the prior distribution $\mathcal{N}(\mu_{\theta,0}, s_{\theta,0}^{-1})$, where the hyperparameters are set to $\mu_{\theta,0}=0, s_{\theta,0}=1.0$.
We assume that the covariance function $C(\cdot,\cdot, \bm{\omega})$ of the Guassian process prior for the nonlinear function $f$ is the RBF kernel
\begin{align}
    C(\bm{x}, \bm{x}'; \omega)=\exp\left(-\frac{\|\bm{x}-\bm{x}'\|^{2}}{2\omega}\right),
\end{align}
with the hyperparameter value $\omega=1.0$.
We construct the set of candidate covariate vectors $\mathcal{X}_{\mathrm{pool}}$ by generating each element from $\mathcal{N}(0, 1.0^{2})$.
The dimension of a covariate vector is $d=10$ and $\mathcal{X}_{\mathrm{pool}}$ contains $400$ sample.
We compare the optimal design, which determines the drawing sample according to (\ref{optimal_design_1}), with the random design, which randomly draws $(\bm{x}_{n+1},t_{n+1})$ from $\mathcal{X}_{\mathrm{pool}}\times \left\{0, 1\right\}$.
The outcome variable $Y_{i}$ is generated according to the model (\ref{eq:model}), where the precision parameter of the noise $s_{\varepsilon}=1.0$.
In both designs, the first 10 samples were randomly selected from $\mathcal{X}_{\mathrm{pool}}\times \left\{0, 1\right\}$, and after that, the samples were selected according to each rule.
The estimator outputs the posterior mean $\mu_{\theta, n}$, and we examined the absolute error $\left|\mu_{\theta,n}-\theta\right|$.
Figure \ref{fig:synthetic} shows the absolute error curves of the Bayes estimator for the random design case and the optimal design case as the functions of the sample size.
The depicted absolute error is the average of 1000 experiments.
We can see that the estimation error of the Bayes estimator in the optimal design case is much smaller than in the random design case, especially when the sample size is small.

\subsection{Experiments on Semi-synthetic Data}
In real-world applications, ground truth causal effects are rarely available.
Thus, we evaluate the optimal design through an experiment on semi-synthetic data.
For the evaluation, we use pre-established benchmarking data for causal inference.
The dataset is constructed from the Infant Health and Development Program (IHDP) \cite{hill2011bayesian}.
Each observation consists of 25 covariates, an indicator variable indicating whether the infant received special care, and an outcome variable representing the final cognitive test score.
The dataset has 747 observations.
Since the dataset also contains counterfactual data, we can calculate the treatment effect for each observation.
We assume that the average of them is the true value of the ATE, and it is $4.021$.
We conducted almost the same experiment as for synthetic data, with a few minor modifications to the experimental conditions.
The hyperparameters are set to $\mu_{\theta,0}=0, s_{\theta,0}=1.0$.
The dot product kernel
\begin{align}
    C(\bm{x}, \bm{x}', \omega)=\omega+\bm{x}^{\top}\bm{x}',
\end{align}
with the hyperparameter value $\omega=1.0$ is used for the Gaussian process prior.
The Bayes optimal estimator requires the value of $s_{\varepsilon}$.
As a rough estimator of $s_{\varepsilon}$, we used the inverse of the sample variance of $\bm{y}$.
As in the previous experiment, We compare the optimal design and random design.
The set of candidate covariate vectors $\mathcal{X}_{\mathrm{pool}}$ is constructed by randomly choosing $400$ observations from the total of $747$ observations.
In both designs, the first 50 samples were randomly selected from $\mathcal{X}_{\mathrm{pool}}$, and after that, the samples were chosen according to each rule.
We examined the absolute error of the posterior mean $|\mu_{\theta,n}-\mathrm{ATE}|$.
Figure \ref{fig: semi-synthetic} shows the absolute error curves of the Bayes estimator for the random case and the optimal design case as the functions of the sample size.
The depicted absolute error is the average of 1000 experiments.
As in the results of synthetic data, we can see that the estimation error of the Bayes estimator in the optimal design case is much smaller than in the random design case, especially when the sample size is small.

\section{Conclusion}
In this paper, we proposed a Bayesian sequential experimental design algorithm tailored to estimate the parametric component of a partially linear model with a Gaussian process prior.
Under some assumptions, the parametric component of a partially linear model can be regarded as a causal parameter, the average treatment effect (ACE) or the average causal effect (ACE).
Therefore, we can say that the proposed method is a sequential experimental design tailored to the estimation of a causal parameter.
In general, causal parameters can be a function of the data generating distribution.
ATE or ACE is one of the causal parameters, and there are various other causal parameters, such as conditional average treatment effect (CATE) or effect modification.
Our study assumed a partially linear model for the data generating mechanism. However, various modeling approaches are possible for the data generating mechanism.
We believe that our proposed method can be applied to other combinations of causal parameters and data-generating mechanisms, which is a subject for future work.
\newpage
\bibliographystyle{unsrt}
\bibliography{ref}

\appendix

\section{Partially Linear Model and Causal Inference}
\label{sec:appendix_a}
In this section, we show that the linear component of a partially linear  model can be regarded as a causal parameter.

In the Neyman-Rubin causal model, the existence of potential outcomes $Y^{(0)}, Y^{(1)}$ is assumed \cite{imbens2015causal}.
It is also assumed that $Y=TY^{(1)}+(1-T)Y^{(0)}$.
The average treatment effect (ATE) is defined as 
\begin{align}
    \mathrm{ATE}=\mathrm{E}[Y^{(1)}]-\mathrm{E}[Y^{(0)}].
\end{align}
When the condition $Y^{(0)}, Y^{(1)}\indep T|\bm{X}$ holds, the treatment assignment is called strongly ignorable.
When the strongly ignorable condition holds,
\begin{align}
    \mathrm{E}[Y^{(1)}]-\mathrm{E}[Y^{(0)}]&=\mathrm{E}_{\bm{X}}[\mathrm{E}[Y^{(1)}|\bm{X}]]-\mathrm{E}_{\bm{X}}[\mathrm{E}[Y^{(0)}|\bm{X}]]\\
    &=\mathrm{E}_{\bm{X}}[\mathrm{E}[Y^{(1)}|\bm{X},T=1]]-\mathrm{E}_{\bm{X}}[\mathrm{E}[Y^{(0)}|\bm{X},T=0]]\\
    &=\mathrm{E}_{\bm{X}}[\mathrm{E}[Y|\bm{X},T=1]]-\mathrm{E}_{\bm{X}}[\mathrm{E}[Y|\bm{X},T=0]].
\end{align}
In our model (\ref{eq:model}), $\mathrm{E}[Y|\bm{X},T]=\theta T+f(\bm{X})$.
Thus, 
\begin{align}
    \mathrm{E}_{\bm{X}}[\mathrm{E}[Y|\bm{X},T=1]]-\mathrm{E}_{\bm{X}}[\mathrm{E}[Y|\bm{X},T=0]]&=\mathrm{E}_{\bm{X}}[\theta+f(\bm{X})]-\mathrm{E}_{\bm{X}}[f(\bm{X})]\\
    &=\theta.
\end{align}

Next, we show that $\theta$ can also be regarded as a causal parameter in Pearl's structural causal inference framework \cite{pearl2000causality}.
We assume that the data is generated according to the following structural equation model.
\begin{align}
    \bm{X}&=\bm{\varepsilon}_{\bm{X}},\\
    Y&=\theta T+f(\bm{X})+\varepsilon_{Y},\qquad \mathrm{E}[\varepsilon_{Y}]=0,\\
    T&=\psi(\bm{X}, \varepsilon_{T}),
\end{align}
where $\psi$ is some nonlinear function, and $\bm{\varepsilon}_{\bm{X}},\varepsilon_{Y}$ and $\varepsilon_{T}$ are independent.
These structural equations specify probability distributions $p(y|t, \bm{x}),p(t|\bm{x})$ and $p(\bm{x})$.
In this case, the distribution of $\bm{X}$ and $Y$ after an intervention $T=t$ is
\begin{align}
    p_{\mathrm{do}(T=t)}(\bm{x},y)=p(y|t,\bm{x})p(\bm{x}).
\end{align}
Marginalizing this distribution with respect to $\bm{X}$, we obtain
\begin{align}
    p_{\mathrm{do}(T=t)}(y)=\int p(y|t,\bm{x})p(\bm{x})\mathrm{d}\bm{x}.
\end{align}
The average causal effect is defined as 
\begin{align}
    \mathrm{ACE}=\mathrm{E}_{\mathrm{do}(T=1)}[Y]-\mathrm{E}_{\mathrm{do}(T=0)}[Y],
\end{align}
where
\begin{align}
    \mathrm{E}_{\mathrm{do}(T=t)}[Y]=\int y\cdot p_{\mathrm{do}(T=t)}(y)\mathrm{d}y.
\end{align}
If the order of integration can be exchanged, $\mathrm{E}_{\mathrm{do}(T=t)}[Y]=\theta t+\mathrm{E}[f(\bm{X})]$, and then $\mathrm{ACE}=\theta$.

\section{Derivation of the Posterior Distribution of $\theta$}
Let $\bm{z}=[z_{1},\ldots,z_{n}]^{\top}=[f(\bm{x}_{1}), \ldots,f(\bm{x}_{n})]^{\top}$.
From the assumption, $\bm{z}$ follows a Gaussian distribution $\mathcal{N}(\bm{0}, \bm{\Phi})$.
The exponent of $p(\bm{y},\bm{z}|\theta,\bm{t},\bm{X})=p(\bm{y}|\bm{z},\theta, \bm{t}, \bm{X})p(\bm{z})$ can be written as
\begin{align}
    -\frac{1}{2}\left(\left[\begin{array}{c}\bm{y}\\ \bm{z}\end{array}\right]-\bm{A}^{-1}\bm{b}\right)^{\top}\bm{A}\left(\left[\begin{array}{c}\bm{y}\\ \bm{z}\end{array}\right]-\bm{A}^{-1}\bm{b}\right)+\mbox{const.},
\end{align}
where
\begin{align}
    \bm{A}&=\left[\begin{array}{cc}s_{\varepsilon}\bm{I}_{n}& -s_{\varepsilon}\bm{I}_{n}\\ -s_{\varepsilon}\bm{I}_{n}& s_{\varepsilon}\bm{I}_{n}+\bm{\Phi}^{-1}\end{array}\right],\\
    \bm{b}&=\left[\begin{array}{c}s_{\varepsilon}\theta\bm{t}\\ -s_{\varepsilon}\theta\bm{t}\end{array}\right].
\end{align}
The inverse matrix of $\bm{A}$ is given by
\begin{align}
    \bm{A}^{-1}=\left[
    \begin{array}{cc}
    s_{\varepsilon}^{-1}\bm{I}_{n}+\bm{\Phi} & \bm{\Phi}\\
    \bm{\Phi} & \bm{\Phi}
    \end{array}
    \right].
\end{align}
Thus, $p(\bm{y}|\theta,\bm{t},\bm{X})$ is a Gaussian with mean $\theta\bm{t}$ and covariance matrix $s_{\varepsilon}^{-1}\bm{I}_{n}+\bm{\Phi}$.
The exponent of $p(\bm{\theta}|\bm{t},\bm{y},\bm{X})\propto p(\bm{y}|\theta, \bm{t}, \bm{X})p(\theta)$ can be written as
\begin{align}
    -\frac{1}{2}\left(\left(s_{\theta}+\bm{t}^{T}(s_{\varepsilon}^{-1}\bm{I}_{n}+\bm{\Phi})^{-1}\bm{t}\right)\theta^{2}-2\left(s_{\theta}\mu_{\theta}+\bm{t}^{T}(s_{\varepsilon}^{-1}\bm{I}_{n}+\bm{\Phi})^{-1}\bm{y}\right)\theta\right)+\mbox{const.}
\end{align}
Thus, $p(\bm{\theta}|\bm{t},\bm{y},\bm{X})$ is a Gaussian with mean $\mu_{\theta,n}$ and variance $s_{\theta,n}^{-1}$ given in (\ref{posterior_mean}) and (\ref{posterior_precision}), respectively.

\end{document}